\documentclass[aps,twocolumn,superscriptaddress,groupedaddress,10pt,nofootinbib]{revtex4} 

\usepackage{graphicx}  	
\usepackage{dcolumn}   	
\usepackage{bm}        	
\usepackage{amssymb}   	
\usepackage{amsmath}
\usepackage{epstopdf}
\usepackage[a4paper,margin=20mm]{geometry}
\usepackage{color}
\newcommand{\eh}[1] {\textcolor{blue}{#1}}

\begin{document}

\title{Stress controlled rheology of dense suspensions using transient flows}

\author{Endao Han}
\email[E-mail:]{endao.han1988@gmail.com}
\affiliation{James Franck Institute, The University of Chicago, Chicago, Illinois 60637, USA}
\affiliation{Department of Physics, The University of Chicago, Chicago, Illinois 60637, USA}
\author{Nicole M. James}
\affiliation{James Franck Institute, The University of Chicago, Chicago, Illinois 60637, USA}
\affiliation{Department of Chemistry, The University of Chicago, Chicago, Illinois 60637, USA}
\author{Heinrich M. Jaeger} 
\affiliation{James Franck Institute, The University of Chicago, Chicago, Illinois 60637, USA}
\affiliation{Department of Physics, The University of Chicago, Chicago, Illinois 60637, USA}

\date{\today}

\begin{abstract}
Dense suspensions of hard particles in a Newtonian liquid can be jammed by shear when the applied stress exceeds a certain threshold. 
However, this jamming transition from a fluid into a solidified state cannot be probed with conventional steady-state rheology because the stress distribution inside the material cannot be controlled with sufficient precision. 
Here we introduce and validate a method that overcomes this obstacle. 
Rapidly propagating shear fronts are generated and used to establish well-controlled local stress conditions that sweep across the material. 
Exploiting such transient flows, we are able to track how a dense suspension approaches its shear jammed state dynamically, and can quantitatively map out the onset stress for solidification in a state diagram. 
\end{abstract}

\maketitle

Suspending solid particles in a liquid creates a more viscous fluid \cite{Batchelor_1977, Maron_Pierce, Krieger_Dougherty}. 
For sufficiently large volume fraction $\phi$ of particles, the suspension becomes non-Newtonian and the viscosity depends on the shearing intensity. Non-Newtonian behaviors commonly include continuous shear thickening (CST) \cite{Wagner_PhysToday, Brady_1985, Cheng_2011}, where the viscosity increases mildly with applied shear and, for larger $\phi$, discontinuous shear thickening (DST) \cite{Barnes_1989, Brown_Review_2014, Brown_JOR}, where the viscosity can increase by more than an order of magnitude. 
Even richer dynamics occur when $\phi$ approaches the threshold for jamming \cite{Nagel_Jamming, SJ}: at sufficiently high shear stress, suspensions can reversibly transform from a viscous fluid into a solidified state \cite{Scott, Peters_Nature}. 
Experiments \cite{Xu_EPL, Fernandez, Boyer} and simulations \cite{Seto, Mari_JOR} have shown that both strong thickening and solidification due to shear are related to a stress-dependent change in particle-particle interactions, which switch from lubrication at low stress to direct, frictional contact at high stress. 
A phenomenological model that unifies CST and DST within a framework based on such stress-dependent interactions was developed by Wyart and Cates \cite{Wyart_Cates}. 
Predictions of this model for the shear thickening regime have been validated by experimental \cite{Poon_Guy, Poon_Hermes, Bonn_PRE} and numerical \cite{Mari_JOR, Ness, Morris_2017} work. 
However, the model also makes predictions for the transition into the shear jammed, solid-like state and these have not yet been tested.

A key reason for this is that conventional rheology experiments, as well as simulations, establish steady-state shearing conditions and assume spatially uniform flows. 
Neither of these is appropriate in situations where a fluid is about to transform into a solid. 
Stress-activated solidification in dense suspensions has been studied extensively under non-stationary conditions. 
This includes impact \cite{Stone, Scott, EHan_NC, Brown_2018_1}, extension \cite{Smith_NC,Sayantan}, or simple shear \cite{Peters_Nature, EHan_PRF}. 
In each case rapid external forcing turns the suspension into a jammed solid, which can ``melt'' and return back to a fluid state once the applied stress is removed \cite{Zhang, vonKann, Scott, Peters_Nature}. 
The key element here is that this dynamic jamming, irrespective of how it is triggered, proceeds via rapidly propagating fronts that spatially concentrate the externally applied shear stress. 
Furthermore, the local stress at the front is controlled by the shearing speed imposed at the boundary \cite{EHan_PRF}. 
Therefore, when the front propagates across the suspension with constant speed, in the co-moving frame it establishes a local environment that is exquisitely stress-controlled.

Here we show how propagating fronts can be exploited to perform stress-controlled rheology in regimes inaccessible to methods based on steady-state driving. 
By generating quasi-one-dimensional fronts in a wide-gap shear geometry, we quantitatively test a major prediction of the Wyart and Cates model \cite{Wyart_Cates} for the location of the boundary delineating DST and shear-jammed states as a function of packing fraction and applied stress.

%
%
\begin{figure}
 \begin{center}
\includegraphics[scale = 1]{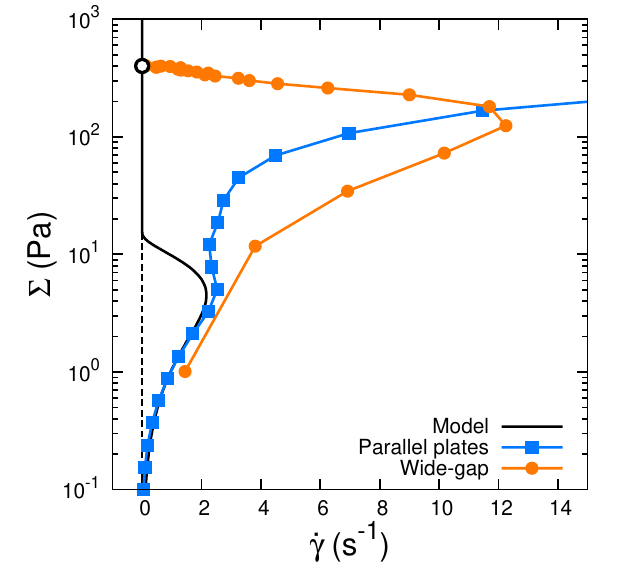}
\end{center}
\caption{\label{Curves} Relation between shear stress $\Sigma$ and shear rate $\dot{\gamma}$ for a suspension in the shear jamming regime, with $\phi = 0.52$. The curves show the prediction of the Wyart-Cates model (solid black line), data obtained from stress-controlled steady-state rheology (blue), and data from wide-gap transient shear flow (orange) at $U_0 = 0.2$~m/s.} 
\end{figure}

Under steady driving conditions, the state of a dense suspension with packing fraction  $\phi$ can be described by two parameters\eh{:} the shear stress $\Sigma$ and the shear rate $\dot{\gamma}$. 
Together, they define a flow curve.  
As an example, the black line in Fig.~\ref{Curves} shows the flow curve predicted by the Wyart-Cates model \cite{Wyart_Cates} for $\phi = 0.52$. 
Such a suspension is well above the frictional jamming packing fraction $\phi_\text{m} = 0.45$ \cite{EHan_PRF} and thus able to shear jam: while $\Sigma$ grows proportionally to $\dot{\gamma}$ at low stress, like a Newtonian fluid, with increasing $\Sigma$ the flow curve bends back towards low $\dot{\gamma}$. 
Eventually, it intersects with the vertical axis where $\dot{\gamma} = 0$~s$^{-1}$. 
At that intersection, the suspension can sustain a non-zero shear stress at zero shear rate, and therefore must have developed a finite, non-zero shear modulus. 
This means the suspension is now jammed.  
We call the stress at this point the onset stress of shear jamming $\Sigma_\text{SJ}$. 
As increasing shear stress is applied, presumably, the jammed suspension remains solid (vertical portion of the black line along the y-axis), until $\Sigma$ eventually exceeds the solid's yield stress (not shown). 
Note that throughout this Letter we plot flow curves in the conventional manner, with $\Sigma$ on the $y$-axis, even though we are discussing stress-controlled protocols where $\Sigma$ is the independent variable. 

When this model flow curve is compared to steady-state experimental data (blue symbols in Fig.~\ref{Curves}), we see obvious deviations. 
The data were taken with a cornstarch suspension of $\phi = 0.52$, using a parallel-plate geometry under stress-controlled shearing conditions. 
The measured $\Sigma$-$\dot{\gamma}$ curve bends only slightly towards low $\dot{\gamma}$. 
It does not keep decreasing and intersect with the $\dot{\gamma} = 0$~s$^{-1}$ axis as expected for shear jamming. 
Instead, with larger $\Sigma$ values the curve bends forward again. 
Such behavior is typical for DST, a regime in which the Wyart-Cartes model predicts s-shaped flow curves \cite{Wyart_Cates,Poon_Guy,Bonn_PRE}. 
The question thus arises: what causes these deviations from the model? 

Rheology experiments can be performed with a variety of geometries. 
The most common geometries are parallel plates, cone and plate, and concentric cylinders (Couette cell). 
For all three geometries, the basic idea is similar: the sample is placed inside a narrow gap (normally 0.1-1~mm) and sheared continuously. 
To obtain the correct viscosity from such measurements, certain conditions must hold: 
the flow must be steady such that $\partial {\bf u} / \partial t = 0$ \eh{;} the shear rate and shear stress in the bulk must have well-defined spatial profiles so they can be calculated from the boundary conditions\eh{;} and there can be no boundary slip. 
For Newtonian fluids, a linear velocity profile across the gap is expected, but for dense suspensions this is not always the case \cite{Xu_EPL, Manneville_2004}. 
When $\Sigma$ exceeds the onset stress of DST, $\Sigma_\text{DST}$, there are complex spatial and temporal rate fluctuations even though the average stress at the boundary is held constant \cite{Poon_Hermes,Manneville_2018,Blair_Fluctuation} and, in addition, boundary slip can be significant \cite{Manneville_2004, Manneville_2018, Peters_Nature, Boersma_JOR}. 
As a result, while it is obviously impossible to maintain a uniformly sheared jammed state under steady-state driving, it is also exceedingly difficult to even approach a jammed state in a truly stress-controlled manner with typical, narrow-gap rheology experiments.

%
%

\begin{figure}
\begin{center}
\includegraphics[scale = 0.9]{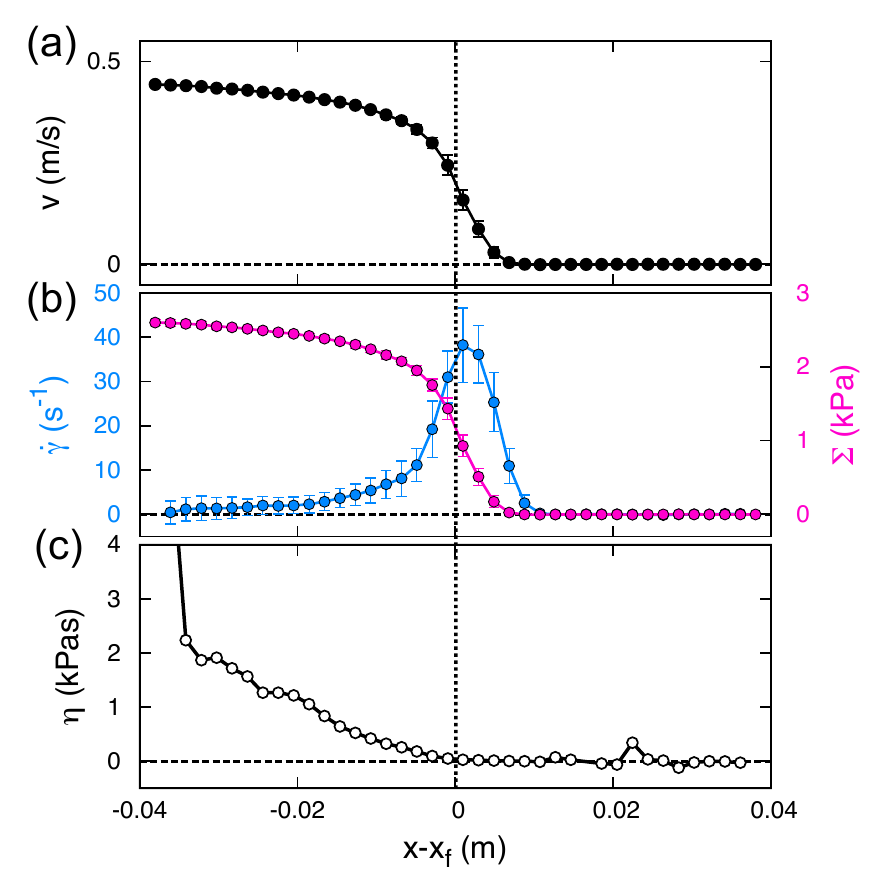}
\end{center}
\caption{\label{ShearThickening} Front profiles in the co-moving frame for a suspension with $\phi = 0.53$. (a) Velocity profile $v(x-x_\text{f})$. (b) Shear stress $\Sigma$ (magenta) and shear rate $\dot\gamma$ (blue) profiles. (c) Local viscosity $\eta = \Sigma / \dot{\gamma}$.   }
\end{figure}

However, there is another, experimentally accessible path to jamming by shear, as shown in Fig.~\ref{Curves}. 
Instead of moving up along the black curve, we can get toward shear-jammed states (e.g., the state indicated by the open black circle) as asymptotic limits of specific $\Sigma$-$\dot{\gamma}$ paths like the one indicated by the orange curve. 
These orange data were taken with the same suspension, but using a wide-gap shear configuration. 
Note how this curve bends back strongly and is able to get much closer to the vertical axis than the blue data.
We now describe how flow curves of this type can be used to perform rheology measurements.

In our wide-gap shear configuration a thin, horizontal layer of suspension floats on a heavy oil, and shear is applied by moving a solid boundary on one side with a constant speed $U_0$ along the y-direction. 
Immediately after start-up, a 1D transient flow  will develop in the x-direction, perpendicular to the movement of the boundary, and spread across the initially quiescent suspension. 
For suspensions with $\phi > \phi_\text{m}$, this flow generates a shear jamming front when $U_0$ is sufficiently fast, i.e., the applied stress sufficiently large  (see \cite{EHan_PRF} for details on the experimental setup and the properties of the shear fronts). 
The profile of such front has an approximately invariant shape, so that 
\begin{equation}
f(x,t) = f(x-U_\text{f}t)
\label{eq:f_X}
\end{equation}
for a front moving toward increasing $x$. 
Here $f$ can be $v$, $\Sigma$, or $\dot{\gamma}$ (see Figs. 2(a) \& (b)), where $v$ is the $y$ component of the local velocity. 
We define the front position $x_\text{f}$ as where $v = 0.45 U_0$, which is also approximately where $\dot{\gamma}$ peaks. 
The front propagates with a constant speed $U_\text{f} \equiv k U_0$, where $k$ is the normalized front propagation speed, which depends on the packing fraction. 
With increasing $U_0$, $k$ reaches a maximum plateau or peak value $k_\text{p}$ that increases with $\phi$ but becomes independent of $U_0$ \cite{EHan_PRF}.

To use the fronts for rheology, we need to know the local shear rate and stress generated by them. 
Given our effectively 1D flow, the equation of motion is 
\begin{equation}
\rho \frac{\partial v}{\partial t} = - \frac{\partial \Sigma}{\partial x}, 
\label{eq:EOM}
\end{equation} 
which reflects the fact that the viscous stress is always balanced by the acceleration of the suspension. 
This allows us to obtain the local shear stress without measuring forces, simply by calculating the stress needed for the suspension to accelerate. 
From Eq.~\ref{eq:f_X} and Eq.~\ref{eq:EOM}, we obtain 
\begin{equation}
\Sigma = \rho U_\text{f} v, 
\label{eq:Sigma}
\end{equation} 
and therefore $\Sigma(x,t)$ has the same shape as $v(x,t)$, but with a prefactor $\rho U_\text{f}$. 
The mean velocity profile $v(x-x_\text{f})$ is shown in Fig.~\ref{ShearThickening}(a). 
Here $v(x,t)$ was shifted by $x_\text{f}$ and then averaged to obtain $v(x-x_\text{f})$. 
The corresponding shear stress is shown in Fig.~\ref{ShearThickening}(b). 
As $(x-U_\text{f}t) \to -\infty$, $v \to U_0$, and $\Sigma$ approaches a constant stress $\Sigma_0 = \rho k U_0^2$. 
Since this stress originates from the acceleration of the whole flow, which develops with little variation in shape before the front reaches a solid boundary, $\Sigma_0$ is very stable.

The local shear rate $\dot{\gamma} = \left| \partial{v}/\partial{x} \right|$ calculated from the averaged velocity profile is also shown in Fig.~\ref{ShearThickening}(b). 
Because ours is a 1D system, we take $\dot{\gamma}$ to be positive for simplicity. 
We can see that both $\Sigma$ and $\dot{\gamma}$ increase at the leading edge of the front ($x > x_\text{f}$). 
However, behind the front ($x < x_\text{f}$), $\Sigma$ keeps increasing and approaches a finite value $\Sigma_0$, while $\dot{\gamma}$ decreases and approaches zero. 
This means that the viscosity $\eta = \Sigma / \dot{\gamma}$ increases dramatically behind the front, as shown in Fig.~\ref{ShearThickening}(c). 
Using our suspension at $\phi = 0.53$ as an example, the solvent viscosity was $\eta_0 = 11 \times 10^{-3}$~Pa\eh{$\cdot$}s, and the suspension viscosity in the quiescent state, prior to shear thickening, was $\eta_\text{N} = 1.3$~Pa\eh{$\cdot$}s. 
At only 2~cm behind the front $x_\text{f}$, the suspension is already almost 1,000-fold more viscous than $\eta_\text{N}$.

Moreover, compared to DST under steady-state conditions, where $\eta \propto \Sigma$, the viscosity increase generated by the shear front is ``beyond discontinuous\eh{,}'' because now $\eta(\Sigma)$ diverges as $\Sigma \to \Sigma_0$ (see Suppl. Mat.). 
In other words, once the front passes, the suspension will evolve toward a solid-like shear-jammed state with finite shear modulus so that $\dot{\gamma} |_{t \to +\infty} \to 0$.

%
%

\begin{figure}
\begin{center}
\includegraphics[scale = 1]{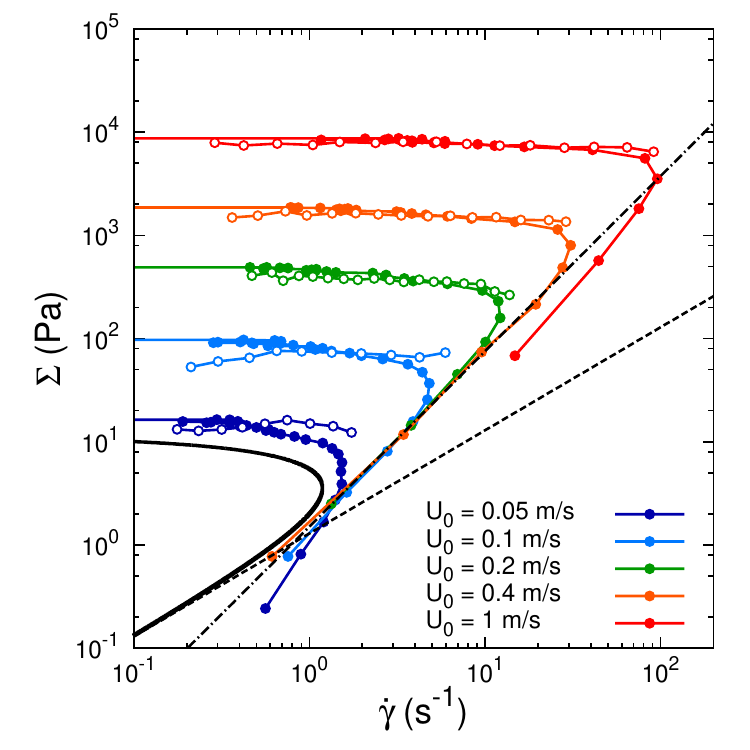}
\end{center}
\caption{\label{Phi532} $\Sigma$-$\dot{\gamma}$ flow curves for a suspension with $\phi = 0.53$, driven at different boundary speeds $U_0$. Solid circles represent data obtained from the velocity profiles as in Fig.~\ref{ShearThickening}. Open circles are obtained by calculating $\dot{\gamma}$ and $\Sigma$ at each time, and then averaging. The solid black curve shows the prediction of the Wyart-Cates model. The dashed black line shows $\Sigma = \eta_\text{N} \dot{\gamma}$. The dash-dot black line is $\Sigma \propto \dot{\gamma}^{1.7}$. }
\end{figure}

While we plot the orange curve in Fig.~\ref{Curves} together with the steady-state prediction and experiment, this transient flow curve needs to be interpreted in a different way. 
For any point on a steady-state flow curve, the overall accumulated strain $\gamma$ is irrelevant since the curve describes a stationary state. 
In contrast, under transient conditions the state of the suspension evolves with time and each point on the $\Sigma$-$\dot{\gamma}$ curve corresponds to a different $\gamma$. 
For the orange curve in Fig.~\ref{Curves} this evolution begins with $\gamma$= 0 and small $\Sigma$ at the start of the shearing process. 
As strain accumulates while the front sweeps through, the suspension experiences increasing shear stress $\Sigma$ and rate $\dot{\gamma}$.
In accordance with Fig. 2(b), $\dot{\gamma}$ then abruptly decreases as the system begins to enter into a jammed state, and $\dot{\gamma}$ vanishes as the flow curve terminates on the vertical axis.

Importantly, the specific stress level within the range of jammed states that is approached in this manner is fully controlled by the shearing speed $U_0$, because as $\dot{\gamma} \to 0$, $\Sigma$ approaches $\Sigma_0 = \rho k U_0^2$, as shown above. 
This means that by changing $U_0$ we can drive the suspension toward different jammed stress levels $\Sigma_0$, as shown in Fig.~\ref{Phi532}. 
The data in this figure were obtained by two analysis methods. 
Solid points are from the mean velocity profiles $v(x-x_\text{f})$, and $\Sigma$ and $\dot{\gamma}$ were extracted the same way as in Fig.~\ref{ShearThickening}. 
Open circles were obtained by first calculating $\Sigma$ and $\dot{\gamma}$ from the velocity profiles at each time step, and then averaging $\Sigma$ and $\dot{\gamma}$. 
The two methods match well, especially at large $U_0$.


\begin{figure}
\begin{center}
\includegraphics[scale = 0.88]{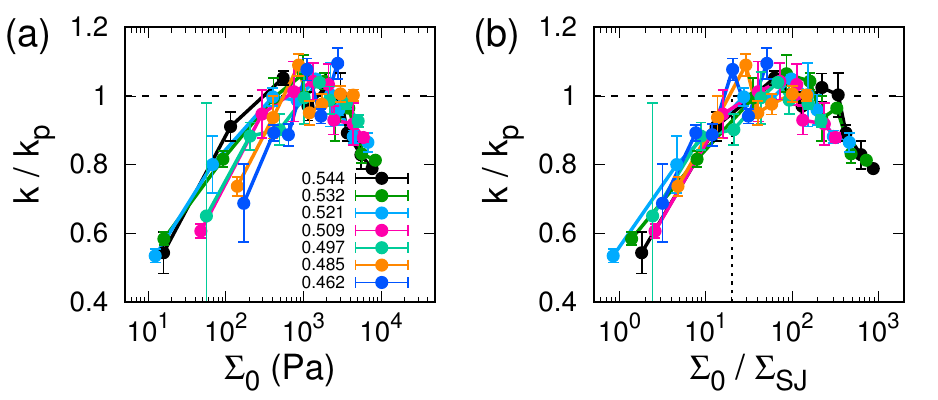}
\end{center}
\caption{\label{Stress_k} Normalized front propagation speed $k / k_\text{p}$ as a function of shear stress $\Sigma$ for different packing fractions $\phi$. (a) $k / k_\text{p}$ as a function of $\Sigma$. (b) $k / k_\text{p}$ as a function of $\Sigma$ normalized by $\Sigma_\text{SJ}$ from Eq.~\ref{eq:Stress_SJ}.}
\end{figure}

For small $U_0$ the stress $\Sigma_0 = \rho k U_0^2$ is no longer directly proportional to $U_0^2$ because the normalized front speed $k$ exhibits a dependence on $U_0$ \cite{EHan_PRF}. 
We see that in Fig.~\ref{Stress_k}(a), where we plot $k$ as a function of $\Sigma_0$, obtained from plots like Fig.~\ref{Phi532}. To compare suspensions prepared with different $\phi$, we normalized $k(\phi)$ by its mean peak height, $k_\text{p}(\phi)$, reached at large $U_0$.
In general, $k/k_\text{p}$ grows from 0 to 1 as $\Sigma_0$ increases, with suspensions at smaller $\phi$ requiring larger stress $\Sigma_0$ to reach the same $k/k_\text{p}$.

If we now plot $\Sigma_0$ relative to the onset stress for shear jamming $\Sigma_\text{SJ}$ predicted by the Wyart-Cates model, we find that the data overlap well (Fig.~\ref{Stress_k}(b)). 
Within the Wyart-Cates model $\Sigma_\text{SJ}$ is given by 
\begin{equation}
\Sigma_\text{SJ} = - \Sigma^* \text{ln} \left( \frac{\phi-\phi_\text{m}}{\phi_0-\phi_\text{m}} \right). 
\label{eq:Stress_SJ}
\end{equation}  
Here $\Sigma^*$ represents the characteristic stress scale, beyond which the dominant interaction between particles changes from lubrication to direct frictional contact, and it is thought to be independent of $\phi$ \cite{Wyart_Cates}. 
For the cornstarch suspensions discussed here, $\Sigma^* = 20.4$~Pa \cite{EHan_PRF}.
This collapse implies that we can identify the predicted shear jamming onset simply by the point at which the $\phi$-dependent front propagation speed $k$ has reached a certain fraction of its maximum value.

From the data collapse in Fig.~\ref{Stress_k}, we can see that when the boundary stress is $\Sigma_0 = \Sigma_\text{SJ}$, the normalized front propagation speed is $k / k_\text{p} \approx 1/2$. 
A front that propagates outward, away from the shearing boundary can be formed in this regime, but its relative propagation speed has not reached the maximum yet. 
This means that though the suspension shear thickens substantially, the final state is still a ``thickened'' state instead of a ``jammed'' state.
As stress increases, $k / k_\text{p}$ approaches and saturates at one. 
The corresponding stress is $\Sigma_0 \approx 20 \Sigma_\text{SJ}$, and this is the stress at which the suspension completely jams in this system.

\begin{figure}
\begin{center}
\includegraphics[scale=1]{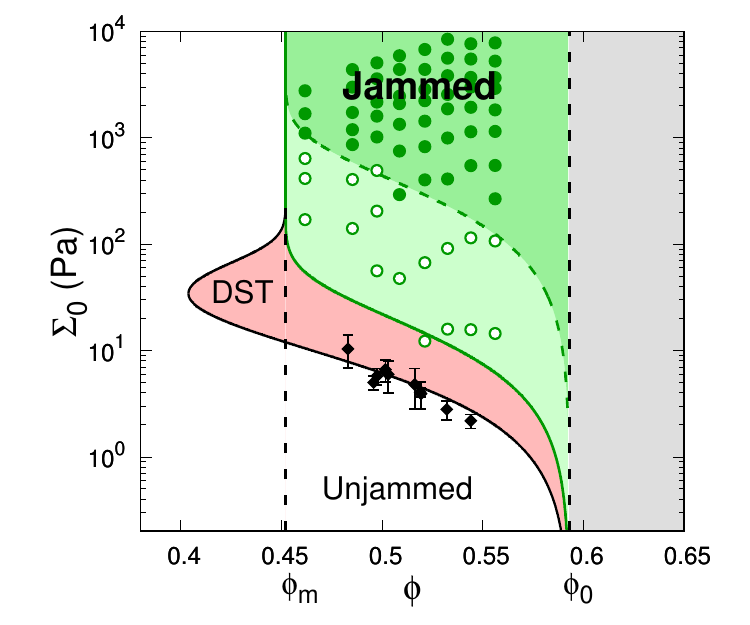}
\end{center}
\caption{\label{StateDiagram} State diagram for cornstarch suspensions. Black diamonds are the onset stress of DST obtained from standard rheology measurements. Green points are obtained from wide-gap transient flows at different $U_0$. Solid points are in the shear jamming regime and open circles are in the transition regime. The red region shows the DST regime predicted by the Wyart-Cates model \cite{Wyart_Cates}.}
\end{figure}

The $k$-$\Sigma_0$ relationship allows us to map out the boundary of the shear jamming regime in the state diagram and compare it with the prediction of the Wyart-Cates model \cite{Wyart_Cates} as shown in Fig.~\ref{StateDiagram}. 
From steady-state rheology experiments, we obtained the frictional and frictionless jamming packing fractions $\phi_\text{m}$ (left) and $\phi_0$ (right) labeled by the vertical, dashed lines. 
Shear jamming is expected only when $\phi$ is between $\phi_\text{m}$ and $\phi_0$ \cite{EHan_PRF}. 
For $\phi < \phi_\text{m}$, when $\Sigma$ increases beyond the upper boundary of the DST region, suspensions are predicted, at least based on model calculations and simulations \cite{Wyart_Cates,Morris_2017}, to return to Newtonian behavior, albeit with larger viscosity.
The solid black curve shows the onset stress $\Sigma_\text{DST}$ for DST, and our experimental measurements (black diamonds) from steady-state rheology with the same suspensions used in the experiments discussed above lie right on top of this boundary. 
This boundary, along with the mechanical properties of suspensions on the left side of $\phi_\text{m}$, have been extensively studied by prior experiments \cite{Poon_Guy, Poon_Hermes} and simulations \cite{Mari, Morris_2017}.

In Fig.~\ref{StateDiagram} the solid green curve is the predicted $\Sigma_\text{SJ} (\phi)$ from Eq.~\ref{eq:Stress_SJ}, which our experiments now can test for the first time quantitatively. 
Measurements are shown by the green data points. 
As discussed above, before the system can completely jam, there is a transition regime between $\Sigma_0 / \Sigma_\text{SJ} = 1$ and the ``full jamming stress'' $\Sigma_0 / \Sigma_\text{SJ} \approx 20$. 
Since $\Sigma_\text{SJ}$ is a function of $\phi$ via Eq. 4,  $\Sigma_0 = 20\Sigma_\text{SJ}$, shown by the dashed green curve in Fig.~\ref{StateDiagram}, also is a function of $\phi$. 
We label the points in the transition regime with open circles, which are the points on the left-hand side of the lowest stress at which $k / k_\text{p} \ge 1$ in Fig.~\ref{Stress_k}. 
All the other points at higher stress are labeled with solid circles in Fig.~\ref{StateDiagram}. 
So instead of going from DST to jamming directly as the stress increases, there is a transition regime where the suspension shows shear thickening ``beyond discontinuous'' ($\eta \propto \Sigma^\beta$, where $\beta \gg 1$, see Suppl. Mat.), but does not jam completely.

In conclusion, we show that conventional steady-state rheology has limitations when testing suspensions in the regime where shear jamming is approached. 
To obtain flow curves for suspensions in this regime, we introduced a new method that takes advantage of transient shear fronts that are easily observable in a wide-gap shear cell. 
As the front propagates, the dense suspension behind the front will evolve towards a shear jammed state, and the stress of this jammed state can be controlled by the speed of the shearing boundary. 
This makes it possible to map out the onset stress of shear jamming for different packing fractions and, for the first time, compare it in a quantitative way to model predictions. 
We find that the boundary of the shear-jammed regime is not sharp, but resembles a smooth cross-over that spans roughly one decade in applied shear stress, corresponding to a factor of two in front propagation speed.

This project originated from a collaboration with Matthieu Wyart. 
We thank Ivo Peters for providing the PIV algorithm, and Abhinendra Singh for helpful discussions. 
This work was supported by the US Army Research Office through grant W911NF-16-1-0078, the Center for Hierarchical Materials Design (CHiMaD), and the UChicago MRSEC (NSF grant DMR-1420709).



\clearpage

\renewcommand{\theequation}{S\arabic{equation}}
\renewcommand{\thefigure}{S\arabic{figure}}
\setcounter{equation}{0}
\setcounter{figure}{0}

{\bf{\Large Supplemental Material}}

\section{Preparation of the suspension samples}

The suspensions used in our experiments were mixtures of cornstarch granules (Ingredion)  dispersed in aqueous solvents. 
The dry cornstarch was stored under conditions of controlled temperature,  $22.5 \pm 0.5^\circ$C, and $44 \pm 2 \%$ relative humidity. 
The solvent was a mixture of caesium chloride (CsCl), glycerol, and deionized water. 
The mass ratio between glycerol and water in the solvent was $65\%:35\%$. 
The density of the solvent was adjusted to $\rho_\text{l} = 1.62 \times 10^3$~kg/m$^3$ in order to match the density of cornstarch particles $\rho_\text{cs} = 1.63 \times 10^3$~kg/m$^3$. 
The viscosity of the solvent was $11 \pm 1$~mPas. 
We prepared the suspension by mixing $m_\text{cs}$ grams of cornstarch particles with $m_\text{l}$ grams of the solvent. 
Then we left it at rest for approximately 2 hours so that the particles were fully wetted and most air bubbles in the suspension had escaped. 
The packing fraction $\phi$ of the suspension is given by \cite{SI_EHan_SOS}
\begin{equation}
    \phi = \frac{1}{1-\psi} \frac{(1-\xi) m_\text{cs}/\rho_\text{cs}}{(1-\xi) m_\text{cs}/\rho_\text{cs}+m_\text{l}/\rho_\text{l}+\xi m_\text{cs}/\rho_\text{w}},
    \label{eq:phi_m_cs}
\end{equation}
where $\rho_\text{w}$ is the density of water, $\xi$ is the mass ratio of moisture in the dry cornstarch particles, and $\psi$ is the porosity of cornstarch particles. 
We used $\xi = 0.13$ and $\psi = 0.31$ in the experiments reported here \cite{SI_EHan_SOS}. \\

\section{$\Sigma$-$\dot{\gamma}$ curves in the shear jamming regime}

\begin{figure*}
\begin{center}
\includegraphics[scale = 1.3]{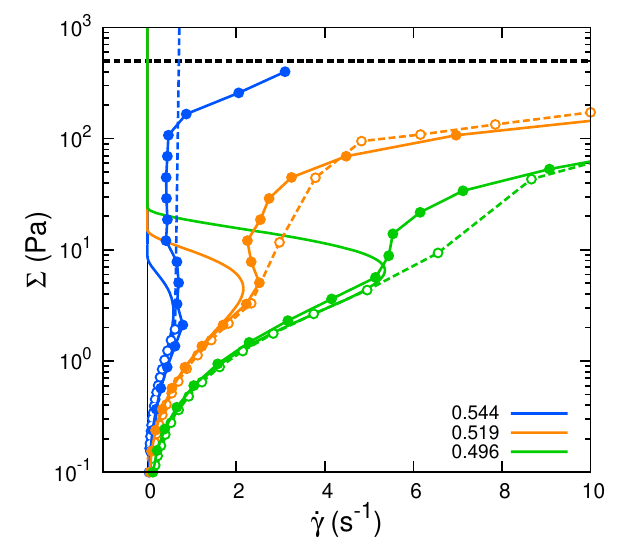}
\end{center}
\caption{\label{SI_SteadyState} $\Sigma$-$\dot{\gamma}$ relations obtained for different suspension packing fractions, using a parallel plate geometry. Shown are predictions of Wyart-Cates model (solid curves), rheology data under stress control (solid circles), and rheology data obtained under rate control (open circles). The black dashed line at $\Sigma = 500$~Pa is the estimated surface tension that confines the liquid-air interface in our rheology experiments. }
\end{figure*}

One obvious difference between the predictions of the Wyart-Cates model and the measurements with narrow-gap rheology is that even when the suspension has a packing fraction in the shear jamming regime ($\phi > \phi_\text{m}$), as $\Sigma$ increases, it does not evolve towards a jammed state with $\dot{\gamma} = 0$~s$^{-1}$. 
The frictional jamming packing fraction $\phi_\text{m}$ \cite{Wyart_Cates} is obtained by fitting the maximum viscosities $\eta_\text{max}$ of the $\eta$-$\dot{\gamma}$ curves at different $\phi$ with a power law function $\eta_\text{max} = \eta_0(1-\phi/\phi_\text{m})^\alpha$, where $\eta_0$ is the solvent viscosity, and the exponent $\alpha$ is a negative number close to -2. 
Details of this process are described in \cite{SI_EHan_PRF}. 
The experimental results at three different packing fractions $\phi = 0.544$, $\phi = 0.519$, and $\phi = 0.496$ are shown in Fig.~\ref{SI_SteadyState}. 
The solid circles are obtained from shear stress controlled experiments, and the open circles are from shear rate controlled experiments. 
All these data are below the upper limit that is provided by the confinement stress $\Sigma_\text{max} \approx 0.1 \Gamma / d \approx 500$~Pa, where $\Gamma$ is the surface tension of the solvent and $d$ is the diameter of the particles \cite{SI_Brown_JOR}. 
Compared to rate-controlled data, stress-controlled $\Sigma$-$\dot{\gamma}$ curves do show an s-shaped section and bend more towards $\dot{\gamma} = 0$~s$^{-1}$; however, none of them actually reach this boundary. 
For all three packing fractions, the suspensions are sufficiently dense  so that they solidify dynamically under impact or rapid onset of shear. 
However, in the steady-state rheology measurements, we only observe DST, or slightly bent s-shaped curves, even in the stress-controlled cases.

\begin{figure*}
\begin{center}
\includegraphics[scale = 1.3]{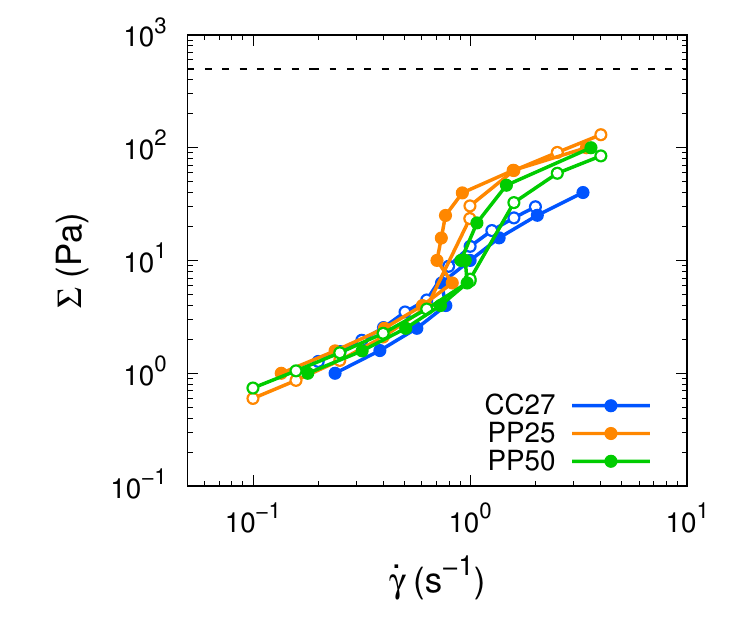}
\end{center}
\caption{\label{SI_Geometry} $\Sigma$-$\dot{\gamma}$ relations measured for the same suspension using three different geometries of the rheometer: concentric cylinders (CC27), 25~mm diameter parallel plates (PP25), and 50~mm diameter parallel plates (PP50). Solid circles are obtained by controlling the applied shear stress and open circles are from shear rate-controlled measurements. The dashed line at $\Sigma = 500$~Pa shows the estimated confinement stress from surface tension. The suspension had a packing fraction $\phi = 0.52$.}
\end{figure*}

The absence of shear jamming in suspensions under steady-state forcing is independent of the geometry of the rheology experiment. 
As shown in Fig.~\ref{SI_Geometry}, we measured the $\Sigma$-$\dot{\gamma}$ curves of the same suspension sample with three different geometries: concentric cylinders (CC27), 25~mm diameter parallel plates (PP25), and 50~mm diameter parallel plates (PP50). Here the brackets designate the specific tools used in our Anton Paar rheometer.
The tests were performed under both shear stress controlled and shear rate controlled conditions. 
The suspension sample was at $\phi = 0.52$ in a 64.9\% (w/w) glycerol/water solution, density matched by adding cesium chloride. 
The viscosity of the suspending solvent was 10.65~mPa$\cdot$s.  
When measured with different shear geometries, the data obtained are not exactly the same, especially in the regime with higher viscosities where $\dot{\gamma} > 1$~s$^{-1}$. 
However, qualitatively they all show a slight s-shaped section that bends back in the stress-controlled experiments, and a discontinuous jump in viscosity in the rate-controlled experiments. 
None of the data indicate steady-state shear jamming. \\

\section{Shear thickening ``beyond discontinuous''}

We can write a power law relation between the viscosity $\eta$ of a fluid and the shear stress $\Sigma$ or shear rate $\dot{\gamma}$ to represent the steepness of the shear thickening transition: 
\begin{equation}
\begin{split}
\eta & \propto \Sigma^\beta, \\
\eta & \propto \dot{\gamma}^{\frac{\beta}{1-\beta}}. 
\label{eq:SR_Power}
\end{split}
\end{equation} 
When $\beta = 0$, $\eta$ is independent of the $\dot{\gamma}$ or $\Sigma$, so the fluid is Newtonian. 
For continuous shear thickening (CST), $0 < \beta < 1$, so $\beta/(1-\beta)$ is positive and finite. 
When discontinuous shear thickening (DST) occurs, $\beta = 1$ or, equivalently, $\beta/(1-\beta) \to \infty$ and consequently the function $\eta(\dot{\gamma})$ jumps vertically at the transition (hence the label ``discontinuous''). 
In the 1D transient flows, the $\Sigma$-$\dot{\gamma}$ curves in Fig.~\ref{Phi532} of the main text show an even more dramatic shear thickening, where $\beta > 1$, and $\beta/(1-\beta) < 0$. We plot the data in Fig.~\ref{Phi532} of the main text in two different ways and present them again in Fig.~\ref{SI_ST}: (a) $\eta$ as function of $\dot{\gamma}$ and (b) $\eta$  as function of $\Sigma$. The suspension shows continuous shear thickening with $0 < \beta < 1$ before it reaches the maximum shear rate. Then, as $\Sigma$ approaches the asymptotic value $\Sigma_0 = \rho U_\text{f} U_0$, $\beta$ grows rapidly and evolves towards infinity. This leads to almost vertical $\eta$-$\Sigma$ curves and slopes of $\beta/(1-\beta) \to -1$ in the $\eta$-$\dot{\gamma}$ relation. The result is a shear thickening transition ``beyond discontinuous'', and shear jamming is obtained when $\beta \to +\infty$. \\

\section{Front propagation speed $k$ and boundary stress $\Sigma_0$ at different $\phi$ and $U_0$}

\begin{figure*}
\begin{center}
\includegraphics[scale = 1]{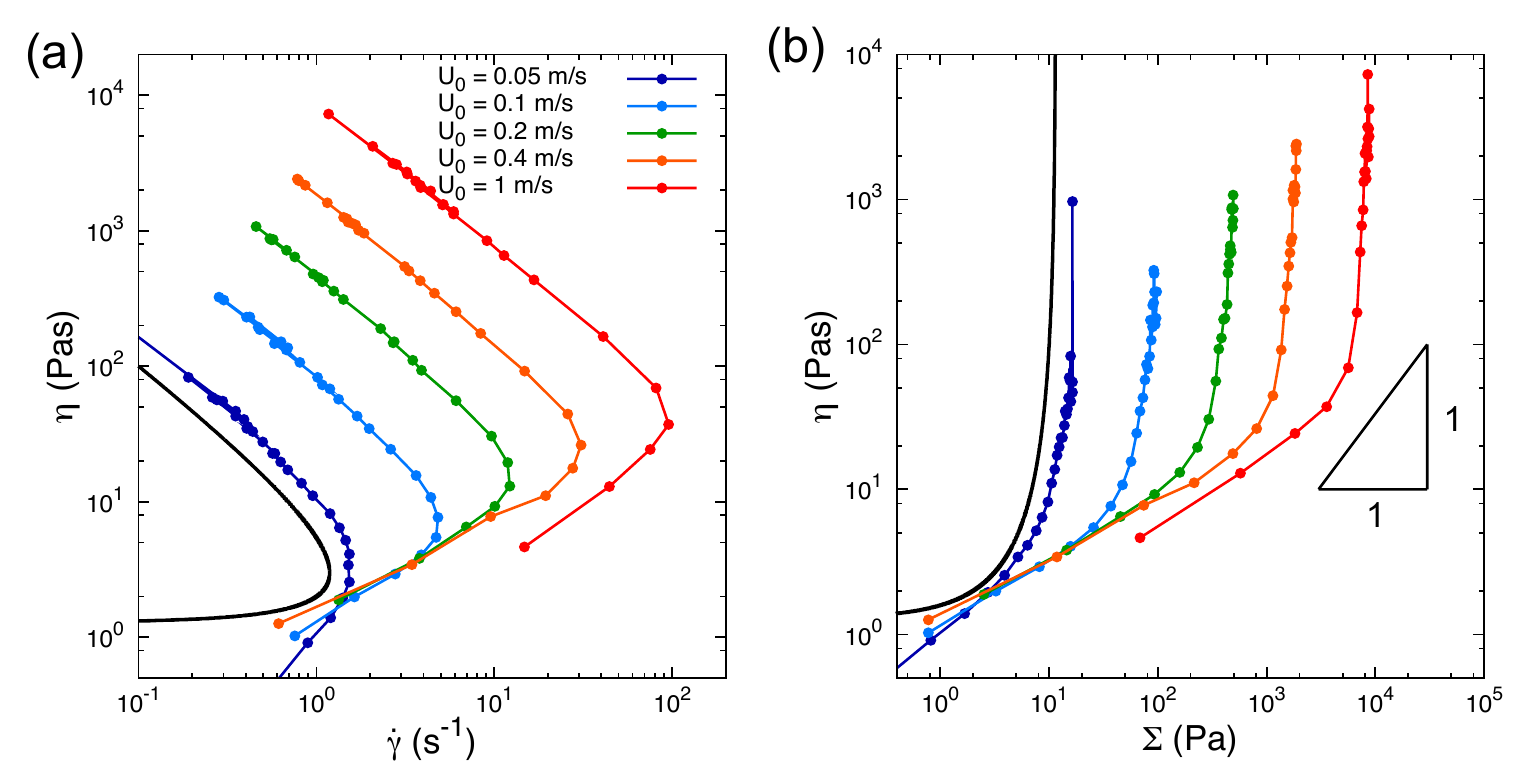}
\end{center}
\caption{\label{SI_ST} Suspension viscosity $\eta$ as function of shear rate $\dot{\gamma}$ and shear stress $\Sigma$. The colors correspond to different boundary speeds $U_0$, as labeled in panel (a). The black curves are predictions of the Wyart-Cates model for the suspension under steady-state forcing. The suspension had a packing fraction $\phi = 0.53$. }
\end{figure*}

\begin{figure*}
\begin{center}
\includegraphics[scale = 1.3]{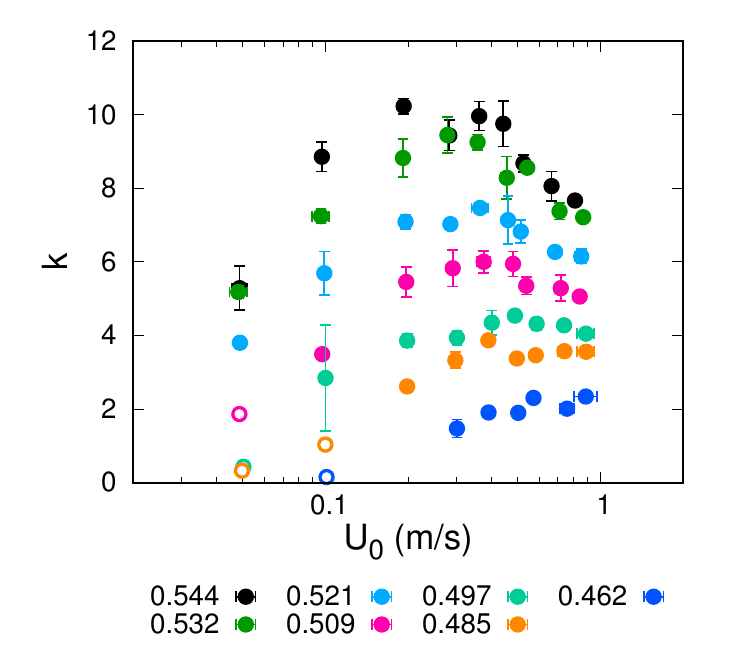}
\end{center}
\caption{\label{SI_K} Dimensionless front propagation speed $k$ as functions of boundary speed $U_0$ for different packing fractions. The solid data points are with $k(\phi) > 0.5 k_\text{p}(\phi)$, and the open circles represent where $k(\phi) \le 0.5 k_\text{p}(\phi)$. The packing fractions are indicated below the plot.}
\end{figure*}

In Fig.~4 of the main text we show how the normalized dimensionless front propagation speed $k / k_\text{p}$ depends on the boundary speed $U_0$. The original data before normalization are shown in Fig.~\ref{SI_K}. 
The transient flow generated by the moving boundary has two different regimes: At slow $U_0$ (low stress), the suspension is in the fluid-like regime, which includes the Newtonian regime, the regime of mild, continuous shear thickening (CST), and the usual DST regime (where $\beta = 1$ in Eq.~\ref{eq:SR_Power}). 
At fast $U_0$ (high stress), the suspension is in the $\beta \to +\infty$ shear thickening regime (as shown in Fig. 3 of the main text and Fig.~\ref{SI_ST} here) or shear jamming regime, where a shear jamming front is generated. 
The fronts we discuss here can only be generated in the high stress regime.

Consider an ideal 1D system with two infinitely large plates both parallel to the $x$-$z$ plane. One of them is at $y = 0$, and the other is at $y \to +\infty$. The suspension is confined between these two plates. For times $t \le 0$, both boundaries are stationary. For times $t > 0$, the boundary conditions become $u(0,t) = U_0$ and $u(H,t) = 0$. In the low $U_0$ limit, the suspension behaves like a viscous Newtonian fluid, and the velocity profile evolves according to  
\begin{equation}
u(y,t) = U_0 \left[ 1-\mathrm{erf}(s/2) \right], 
\label{eq:NF}
\end{equation} 
where $s = y/(\sqrt{\nu t})$ is a characteristic length scale and $\text{erf}(x)$ is the error function \cite{SI_Fluid}. 
Here $\nu = \eta / \rho$ is the kinematic viscosity of the suspension in the Newtonian or mildly shear thickening regimes. 
In the regime of $\beta < 1$ (Eq.~\ref{eq:SR_Power}), the flow profile will be more like that described by Eq.~\ref{eq:NF}. 
However, to generate flow profiles as shown in Fig.~2 of the main text, the shear rate $\dot{\gamma}$ must decrease as the shear stress $\Sigma$ increases, so the $\Sigma$-$\dot{\gamma}$ curves must bend back like those in Fig.~3 of the main text.

\begin{figure*}
\begin{center}
\includegraphics[scale = 1.3]{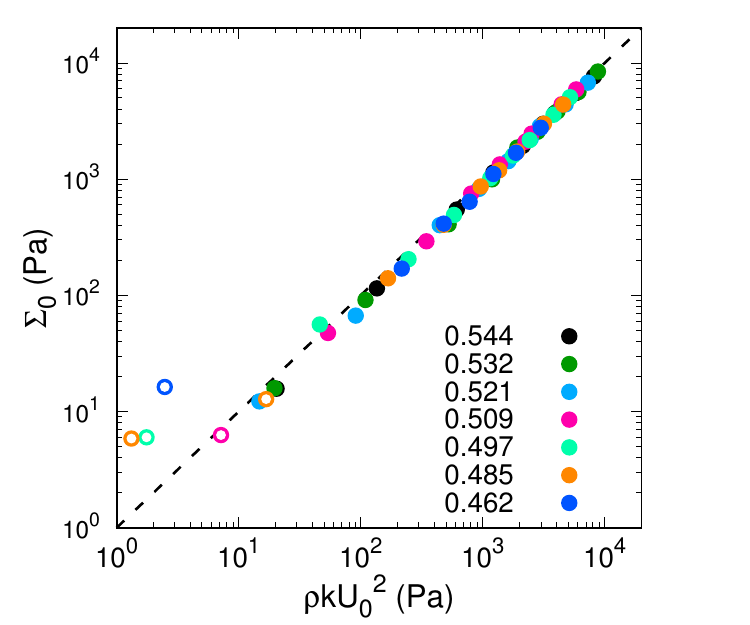}
\end{center}
\caption{\label{SI_Stress_kU2} Comparison between the boundary stress $\Sigma_0$ measured from acceleration and the calculated boundary stress based on k and $U_0$. The color labels are the same as in Fig.~\ref{SI_K}. The dashed black line shows $\Sigma_0 = \rho k U_0^2$. }
\end{figure*}

In the low-stress, fluid-like regime, the flow does not exhibit the two properties required of fronts: invariant shape and constant propagation speed. As a result, $\Sigma_0 = \rho k U_0^2$ does not apply. In Fig.~\ref{SI_Stress_kU2}, the boundary stress $\Sigma_0$ obtained by integrating acceleration (Eq.~2 of the main text) is directly compared with $\rho k U_0^2$. We can see that for points in the high stress regime, where fronts are generated, the relation works well, while in the low stress regime, there are apparent deviations. To separate the two regimes in drive velocity, we take $k = 0.5k_\text{p}$ as a safe threshold. This corresponds to a stress boundary for $\Sigma_0$ that agrees very well with the shear jamming onset stress $\Sigma_\text{SJ}$ predicted by Wyart and Cates.

\section{Comparison of the State diagram with that in Ref.~13}

\begin{figure*}
\begin{center}
\includegraphics[scale=1.3]{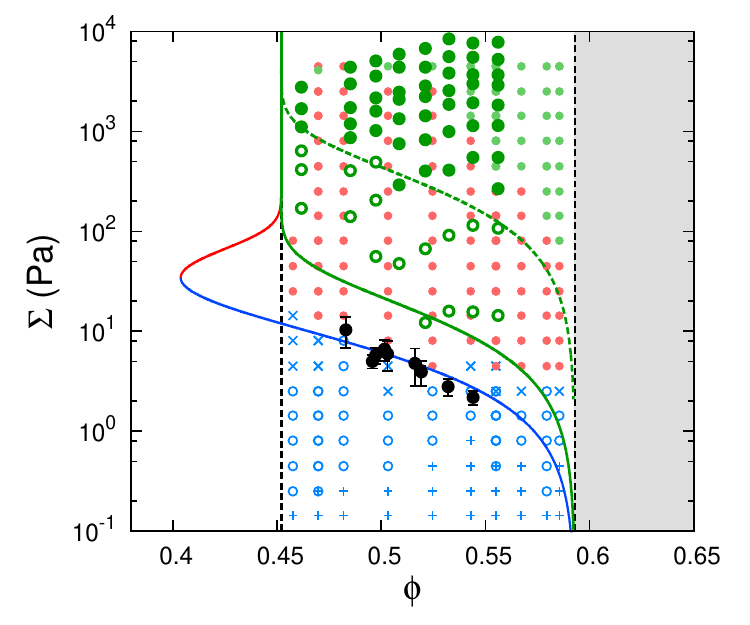}
\end{center}
\caption{\label{SI_StateDiagram_OLD} State diagram for cornstarch suspensions. The larger, darker green points are from wide-gap transient flows at different $U_0$ and are the same as in Fig.~5 of the main text. Among them, the filled circles correspond to the shear jamming regime, and the open circles to the transition regime. The boundaries and the black data points are identical to Fig.~5 of the main text. The smaller, lighter colored data points are from Ref.~13 of the main text. The symbols indicate shear thinning (light blue +), Newtonian (light blue $\circ$) behavior, CST (light blue x), DST (light red filled circle), and shear jamming (light green filled circle). }
\end{figure*}

In Ref.~13 of the main text, Peters {\it et al.} mapped out a state diagram for dense suspensions with a wide-gap Couette cell. 
The stress boundary of shear jamming was found by dropping a small metal sphere onto the surface of a suspension and testing for sinking, while the suspension was continually sheared between the concentric cylinders. 
If the sphere did not sink below the surface, the suspension was labeled jammed; if the sphere sank, the suspension was labeled as not jammed. 
The results are shown in Fig.~\ref{SI_StateDiagram_OLD} and compared directly to results obtained with the method described in the main text. 
The light colored data points are from the experiments of Peters {\it et al}.
The onset stress for shear jamming obtained from their measurements is the boundary between the light red (DST) and light green (shear jammed) solid circles.
Compared to Ref.~13, our new method provides a more accurate determination of the stress boundary of shear jamming, especially in the low $\phi$ regime. 
Another major difference is that in Peters {\it et al.}, the vast red region below shear jamming was identified as DST regime, because data were taken after the flow reached a steady state. 
In that case, an $\eta$-$\Sigma$ curve with slope larger than unity is not likely to be observed. In the same regime the method described here, by contrast, can measure larger slopes straightforwardly.

\section{Correlation between strain $\gamma$ and stress $\Sigma$}

In our effectively 1D shear geometry, the jamming front has a specific velocity profile as shown in Fig.~2. In a transient flow like this, the local shear stress $\Sigma$ and the accumulated strain $\gamma$ are always correlated. From Eq.~\ref{eq:Sigma} we already know that the local shear stress and velocity have the same profiles. This also applies to $\gamma$. At any position $x$ and time $t$, we have
\begin{equation}
\gamma(x,t) = \int_0^t \dot{\gamma}(x,t') dt' = -\int_0^t \frac{\partial v(x,t')}{\partial x} dt'. 
\label{eq:SI_gamma}
\end{equation}
In the following, we take $\gamma$ and $\dot{\gamma}$ to be positive to keep things simple. From Eq.~\ref{eq:f_X}, we get $\partial f / \partial x = -\partial f / (U_\text{f} \partial t)$. Replacing $\partial v / \partial x$ with $-\partial v / (U_\text{f} \partial t)$ in Eq.~\ref{eq:SI_gamma}, we obtain
\begin{equation}
\gamma(x,t) = \frac{v(x,t)}{U_\text{f}}. 
\label{eq:SI_gamma_2}
\end{equation}
Replacing $v(x,t)$ in Eq.~\ref{eq:Sigma} with Eq.~\ref{eq:SI_gamma_2}, we arrive at 
\begin{equation}
\Sigma(x,t) = \rho U_\text{f}^2 \gamma(x,t). 
\label{eq:SI_SS}
\end{equation}
This means that for such fronts, in the shear jamming regime, the local stress is always proportional to the local accumulated strain: As $\gamma$ approaches its asymptotic value $\gamma_\infty$, $\Sigma$ approaches its asymptotic limit $\Sigma_0$ in the same way.

\section{$\Sigma$-$\dot{\gamma}$ curves for different packing fractions $\phi$}

\begin{figure*}
\begin{center}
\includegraphics[scale = 1]{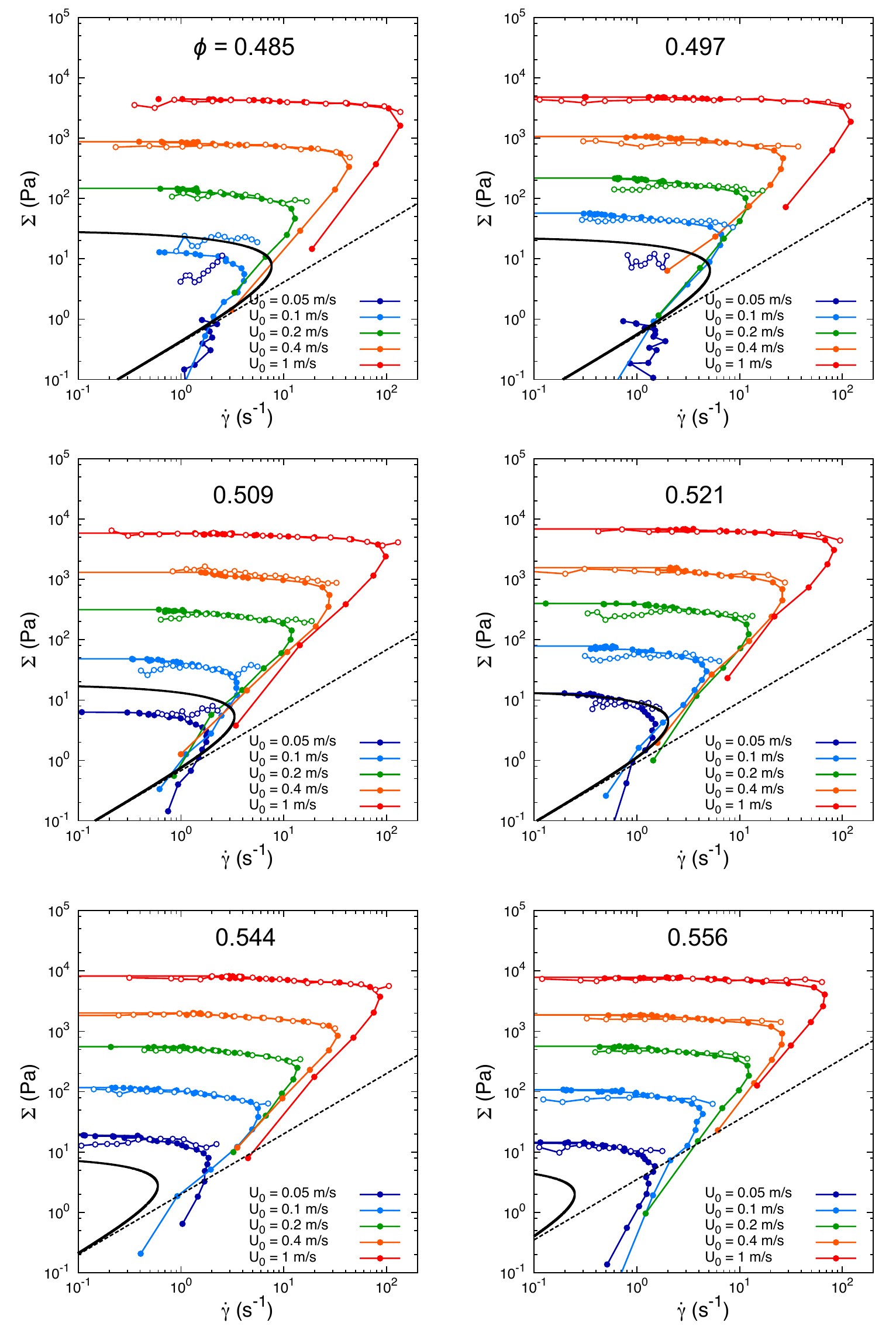}
\end{center}
\caption{\label{SI_SR} $\Sigma$-$\dot{\gamma}$ curves for different packing fractions $\phi$. For each packing fraction, data for five shear speeds $U_0$ are shown. In each panel, the solid black curve indicates the prediction of the Wyart-Cates model at the corresponding $\phi$. The dashed black line shows $\Sigma = \eta_\text{N} \dot{\gamma}$, where $\eta_\text{N}$ is the viscosity of the suspension in the Newtonian regime. }
\end{figure*}

To demonstrate how the states of suspensions at different $\phi$ evolve when driven by different $U_0$, we plot  in Fig.~\ref{SI_SR} the experimentally obtained $\Sigma$-$\dot{\gamma}$ curves for seven different packing fractions. All packing fractions shown here are above $\phi_\text{m}$. The labeling is the same as used for Fig.~\ref{Phi532} in the main text. As mentioned above, at low packing fractions, for example $\phi = 0.485$ and $\phi = 0.497$, shear jamming fronts do not form slow speeds, such as $U_0 = 0.05$~m/s. In this regime the flow profiles do not have invariant shapes and thus the shifting and averaging method we used to obtain Fig.~2 of the main text does not apply. As a result, we do not show the solid data points here.


\clearpage

\end{document}